\documentclass[%
preprint,
 amsmath,amssymb,
 aps,
]{revtex4-2}

\usepackage{graphicx}% Include figure files
\usepackage{dcolumn}% Align table columns on decimal point
\usepackage{bm}% bold math

\usepackage{xcolor}
\usepackage{hyperref}
\usepackage{comment}
\newcommand{\CXP}[1]{\textcolor{black}{#1}}
\newcommand{\MDGrevise}[1]{\textcolor{black}{#1}}

\newcommand{\Ca}{\mathit{Ca}}

\newcommand{\agg}{\textit{agg}}
\newcommand{\gammadot}{\dot{\gamma}}

\usepackage[english]{babel}

\begin{document} 

% \preprint{APS/123-QED}

\title{Microcirculatory blood flow with aberrant levels of red blood cell aggregation}% Force line breaks with \\

\author{Xiaopo Cheng$^1$}

\author{Dell Zimmerman$^1$}

\author{Elizabeth Iffrig$^{2,3}$}

\author{Wilbur A. Lam$^{2,3}$}
\email{wilbur.lam@emory.edu}

\author{Michael D. Graham$^1$}
\email{mdgraham@wisc.edu}

\affiliation{$^1$Department of Chemical and Biological Engineering, University of Wisconsin-Madison, Madison, WI 53706, USA \\
$^2$Aflac Cancer and Blood Disorders Center of Children's Healthcare of Atlanta, Department of Pediatrics, Emory University School of Medicine, Atlanta, GA 30307, USA \\
$^3$Wallace H. Coulter Department of Biomedical Engineering, Georgia Institute of Technology and Emory University, Atlanta, GA 30332, USA
}

\date{\today}

\begin{abstract}
Recent clinical results indicate that aberrant erythrocyte aggregation in hematological disorders is accompanied by endothelial damage and glycocalyx disruption, but the underlying biophysical mechanisms remain unclear. This study uses direct computational modeling to explore how red blood cell (RBC) aggregation impacts shear stress in small blood vessels, highlighting the increased risk of vascular damage. RBC aggregation creates a heterogeneous distribution, leading to variations in the cell-free layer thickness and fluctuating wall shear stress, especially near vessel walls. This effect aligns with experimental findings on endothelial disruption linked to RBC clustering near the wall, potentially reducing the protective glycocalyx layer. The power spectral density analysis of wall shear stress fluctuations reveals that, with RBC aggregation, there is a distinct peak near frequency \( f = 0.04 \), indicating increased fluctuations due to aggregated RBC clusters traveling close to the vessel wall. The presence of aberrant cells in blood disorders, modeled here by sickle cells, further amplifies these effects, as aggregation-enhanced margination drives sickle cells closer to vessel walls, exacerbating shear stress fluctuations and increasing the likelihood of vascular injury and inflammation. Simulations show that curved vascular geometry, with curvature accentuating RBC clustering near vessel walls, intensifies aggregation-induced wall shear stress fluctuations and increases the risk of vascular damage, particularly in sickle cell disease where sickle cells marginate closer to the wall.
\end{abstract}

\maketitle

\section{Introduction and background}

 The tendency of red blood cells to aggregate plays a significant role in blood flow and circulation. The primary cause of RBC aggregation is the interaction between certain proteins on the surface of red blood cells and various plasma proteins, particularly fibrinogen. Fibrinogen, a glycoprotein found in the plasma, acts as a bridge between adjacent red blood cells, promoting their adhesion and aggregation. Other factors, such as cell surface charges and the presence of certain substances in the plasma, also contribute to the degree of aggregation. While some degree of red blood cell aggregation is normal, aberrant aggregation levels can have implications for blood flow and cardiovascular health. 
An important effect that is observed in a variety of disorders is that endothelial cells lining the blood vessels are dysfunctional and in a pro-inflammatory state. This is associated, for example in COVID, sepsis, and sickle cell disease, with substantially increased risk of stroke and other serious circulatory problems \cite{Druzak.2023.10.1038/s41467-023-37269-3,Ince.2016.10.1097/shk.0000000000000473,Nader:2020bc,hahn2009mechanotransduction, walther2021mechanotransduction, wang2016flow, he2020endothelial, panciera2017mechanobiology}.
Very recently, using a microfluidic microvasculature model, Druzak et al.~ \cite{Druzak.2023.10.1038/s41467-023-37269-3} established for the first time that flow of blood with a high level of fibrinogen (and thus propensity to aggregate) directly damages the glycocalyx, the layer of glycoproteins covering cell membranes of endothelial cells. Understanding the mechanisms and factors influencing red blood cell aggregation is essential for gaining insights into vascular health and developing strategies for managing conditions related to blood flow and circulation. 
{This study uses cell-level simulations to investigate erythrocyte aggregation in the microcirculation and its interactions with the vessel wall, considering factors such as blood disorders, flow conditions, and cellular physical properties.}

\subsection{Response of the vascular endothelium to stress}
Endothelial cells mechanotransduce the shear forces of the hemodynamic microenvironment into cellular biological signals \cite{harrison2006endothelial,Chien2007,abe2014novel}. Bao \emph{et al.} \cite{Bao1999} measured the mRNA expression of MCP-1 and PDGF-A, two genes related to endothelial inflammation, by endothelial cells subjected to laminar flows with different well-defined wall shear stress profiles, namely (1) ramp flow in which shear stress was smoothly changed from zero to a maximum value and then sustained, (2) step flow in which shear stress was abruptly applied at flow onset followed by steady shear stress for a sustained period, and (3) impulse flow in which shear stress was abruptly applied and then removed. A steady shear stress profile was also considered for comparison. They showed that rapid changes in shear stress and steady shear stress, respectively, stimulate and diminish the expression of both genes. In particular, the impulse flow case was found to induce the highest level of endothelial expression of the pro-inflammatory signals. Other studies \cite{Davies1997,DeKeulenaer1998,Chen1999Mechano,Akimoto2000,Davis2001,Dekker2002,Hsiai2003,Sorescu2004,Li2005Molecular,harrison2006endothelial} have also elucidated a variety of shear stress-induced signal transduction pathways of the endothelium, and revealed the opposite effects of steady and unsteady shear stresses on endothelial dysfunction. 

To investigate the role of cellular interaction with the vascular endothelium in hematological diseases, Caruso et al. \cite{caruso2019stiff} developed an \textit{in vitro} microvasculature model comprised of endothelial cells cultured on the inner surface of a microfluidic system. Sickle cell disease (SCD) RBCs were spiked into normal RBC suspension, suspended, and then perfused into this endothelialized microfluidic. They found that VCAM1, a biomarker of endothelial cell dysfunction, was upregulated when exposed to flowing SCD RBCs. These results together imply that purely physical interactions between endothelial cells and SCD RBCs are sufficient to cause endothelial inflammation.  A similar study was performed to model iron deficiency anemia (IDA)~\cite{Caruso.2022.10.1016/j.isci.2022.104606}. IDA suffers have a subpopulation of RBCs that are smaller and stiffer than healthy RBCs, and this study found, as with SCD, that perfusion of mixtures of healthy and iron-deficient RBCs leads to upregulation of markers of pro-inflammatory response.  Furthermore, theoretical and computational studies indicate that diseased cells are likely to marginate to blood vessels walls because of their smaller size and higher stiffness \cite{Kumar:2012ie,HenriquezRivera:2015fx,HenriquezRivera:2016wb}, and that margination leads to increased frequency of events of high wall shear stress \cite{Zhang:2020jt,Caruso.2021.10.1182/blood-2021-150591,Caruso.2022.10.1016/j.isci.2022.104606,Cheng.2023.10.1126/sciadv.adj6423}. These observations indicate that marginated aberrant cells may play a substntial role in the vasculopathy observed in many hematological disorders.

\vspace{-0.1in}
\subsection{Aggregation in blood flow: experimental and computational observations}
Aggregation of RBCs at rest and low shear rates is a well-known and well-studied phenomenon. Microstructurally, it leads, in combination with the discoidal shape of RBCs to the formation of stacks of RBCs known as rouleaux. Rheologically, it leads to a small yield stress and a regime at low shear rates ($\lesssim 100 s^{-1}$) of substantial shear-thinning and thixotropy \cite{Armstrong.2022.10.1122/8.0000346}.  
% The blood literature is vast and here we only review studies with particular relevance to the proposed work.

Mechanistically, aggregation of RBCs is generally attributed to the presence of fibrinogen, which binds to RBCs membranes leading to aggregation by a bridging mechanism. However, classical depletion
	flocculation (which is osmotic in origin \cite{Russel:1989vm}) due to the presence of
	the large number of albumin and globulin molecules also plays some role \cite{Neu:2002uc,Rampling:2004uq}. Indeed, aggregation is frequently induced in experiments with blood via the depletion mechanism by addition of dextran. At the level of modeling undertaken in the present work, we will not need to distinguish between the two mechanisms, simply using a phenomenological potential to describe aggregation forces between points on neighboring cells.  
We detail the approach in Section \ref{sec:formulation}.
	
One natural question regarding aggregation is the effects it might have on how cells are distributed in the microcirculation (the blood vessels $< 100 \mu m$ in diameter) during flow. Key features of this distribution are the presence of (1) the cell-free layer (CFL),  and (2) the phenomenon of margination in which white cells and platelets (and also aberrant cells, as noted above) are preferentially found near blood vessel walls, i.e.~in the CFL.  The basic mechanism of these phenomena in terms of cell-cell collision dynamics and wall-induced migration is understood \cite{Kumar:2012ie,HenriquezRivera:2015fx}.  Regarding the cell-free layer, one might expect that this is thickened by aggregation as the RBCs tend to pack together under attractive forces. Several experimental studies have confirmed this intuition (as do simulation results below).  Ong et al. \cite{ong2010effect} investigated the impact of red blood cell aggregation and flow rate on the arteriolar cell-free layer in the rat cremaster muscle. By inducing aggregation with dextran 500 infusion, they observed enhanced cell-free layer formation in arterioles. Interestingly, the same group \cite{ong2011cell} also later found that the \emph{variance} of the CFL thickness also increased under aberrant aggregation conditions -- we also see this in our results below. Similar experimental observations were made using a microchannel with a T-junction \cite{sherwood2012effect}.

Claveria et al \cite{claveria2021vitro} studied the distribution and margination of sickle red blood cells (sRBCs).  In nonaggregating suspending fluid, low-density (more flexible) sRBC stayed at the channel center, while the densest (least flexible) cells segregated towards the walls. Our prior computational studies, described in  refs.~\cite{Zhang:2020jt,Cheng.2023.10.1126/sciadv.adj6423}, capture this effect. As in the studies noted above, aggregation led to an increase in CFL thickness, while also inhibiting segregation. The latter is a somewhat counterintuitive result, as a thicker CFL leaves more room near blood vessel walls for stiff or small cells to reside, promoting margination. Again, the computational results below are consistent with these experimental observations.

 In COVID, fibrinogen levels can be several times normal, resulting not only in increased aggregation (hypercoagulability) but increased plasma viscosity (hyperviscosity), again up to several times normal \cite{Truong.2021.10.1111/trf.16218}. Thus the overall blood viscosity is also much higher than normal, despite the hematocrit generally being lower \cite{Nader.2022.10.1002/ajh.26440}. This is a distinctive characteristic of COVID; in other hyperviscosity syndromes high viscosity comes from immunoglobulin association, not fibrinogen.

{Recent advances have provided valuable insights into erythrocyte aggregation. Dynar et al. \cite{dynar2024platelet} investigated platelet margination under varying RBC aggregation energies using 2D simulations. Their findings indicate that moderate RBC aggregation enhances platelet margination in microcirculation, whereas excessive aggregation leads to a decline in margination. Using a combination of experiments and simulations, Dasanna et al. \cite{dasanna2022erythrocyte} examined the dependence of RBC sedimentation rates on fibrinogen concentration and its relationship to the microstructure of gel-like erythrocyte suspensions. Their results demonstrated that higher fibrinogen concentrations strengthen intercellular interactions, creating larger voids in the gel structure, which in turn increases permeability and accelerates sedimentation. Furthermore, recent research \cite{puthumana2024dissociation} revealed that aggregate dissociation rates increase sharply upon reaching a critical extension rate, which falls within the range of microcirculatory conditions. This suggests significant variability in aggregate sizes in vivo.
}

Many computational studies have addressed the role of RBC aggregation in various aspects of blood flow.  Liu et al. \cite{liu2004coupling, liu2006rheology} proposed a 3D model to explore RBC aggregation and its impact on blood rheology. Their approach effectively captures three significant phenomena in blood rheology: shear rate-dependent viscosity, the influence of cell rigidity on viscosity, and the Fahraeus-Lindqvist effect. Bagchi et al. \cite{bagchi2005computational} presented a 2D simulation of RBC aggregation using an immersed boundary method, considering rheological properties and cell-cell adhesion kinetics. Their results show that higher cytoplasmic viscosity and membrane rigidity lead to stable aggregate formation.  Zhang et al. \cite{Zhang:2008kr} used an immersed boundary lattice Boltzmann algorithm to simulate RBC behavior in shear flows. In 2D shear flows, RBCs aggregate into rouleau-like structures influenced by interaction strength, exhibiting concave shapes in weak interactions and convex shapes in strong interactions. These aggregates rotate or separate in shear flows, aligning with experimental observations. Fedosov et al. \cite{fedosov2011predicting} accurately predict blood viscosity dependence on shear rate and hematocrit. Xu et al. \cite{xu2013large} examined the flow of $49,512$ RBCs at a $45 \%$ concentration under aggregating forces. Results align well with experimental findings, revealing uniform RBC distributions at moderate shear rates $(60-100 / \mathrm{s})$ and significant aggregation structures at a lower shear rate $\sim 10 / \mathrm{s}$. Yoon et al.\cite{yoon2016continuum} developed a method for modeling inter-cellular interactions among multiple adjacent RBCs, analytically determining interaction energy. This approach differs from previous curve-fitting methods for aggregation, enabling predictions of various effects of physical parameters on depletion and electrostatic energy among RBCs. Deng et al. \cite{deng2020quantifying} used patient-specific computational simulations to explore fibrinogen-dependent aggregation dynamics in T2DM RBCs, integrating in vitro experiments. The study sheds light on complex cell-cell interactions, emphasizing the significance of RBC aggregation and disaggregation dynamics in patients at a higher risk of microvascular complications. 

Despite the substantial number of computational studies of RBC aggregation, they do not consider disease states, and accordingly none have focused on the key issues that we address in the present work: (1) abnormally high levels of aggregation, which will affect blood flow at high strain rates, (2) dynamic and spatially inhomogeneous RBC distributions in the presence of aberrant aggregation, (3) the wall shear stress (endothelial) environment.

{In this study, RBC aggregation in the microcirculation was investigated through cell-level simulations, with a focus on its effects on cell distribution, cell-free layer dynamics, and cell-wall interactions. Various influencing factors were considered, including cellular physical properties, flow conditions, vessel geometry, and blood disorders. Results indicate that RBC aggregation increases the average thickness of the cell-free layer due to mutual cell attraction, but also introduces substantial fluctuations in the cell-free layer, with more instances of erythrocytes clustering very close to the vessel wall. These cell clusters induce significant wall shear stress fluctuations, potentially damaging the glycocalyx layer. In cases involving blood disorders, where a small fraction of cells exhibit increased rigidity, erythrocyte aggregation enhances the segregation of various cellular components, while the margination of stiff cells and RBC aggregation together intensify wall stress fluctuations. Additionally, in curved vessels, larger cell-free void areas were observed in aggregation, suggesting that vessel geometry promotes cell clustering and contributes to spatial heterogeneity in the cell-free layer.}

\section{FORMULATION}\label{sec:formulation}

We simulate a suspension of RBCs, modeled as deformable fluid-filled elastic capsules, flowing through rigid straight and curved cylindrical tubes. For blood disorders such as sickle cell disease, RBC suspensions are modeled as binary mixtures of normal and sickle RBCs. In these binary suspensions, the number fraction consists of 0.9 normal RBCs and 0.1 aberrant sickle RBCs. A suspension of normal RBCs without intercellular aggregation is considered a control. For most simulations here, the overall volume fraction (hematocrit), $\phi$, of blood cells is around $20 \%$, consistent with the hematocrit in the microcirculation \CXP{\cite{Klitzman.1979.10.1152/ajpheart.1979.237.4.h481, Sarelius.1982.10.1152/ajpheart.1982.243.6.h1018}}. The suspending fluid, blood plasma, is considered incompressible and Newtonian with a viscosity of about $\eta=1.10-1.35 \mathrm{mPas}$. 
{In this study, we assume for computational reasons a viscosity ratio of 1 between the intercellular matrix and plasma. While typical RBCs may exhibit viscosity ratios of up to 15 \cite{Recktenwald.2022.10.1016/j.bpj.2021.12.009}, aberrant RBCs associated with blood disorders can display even higher values. Nevertheless, previous research has shown that the dynamics of both individual cells \cite{Zhang:2019cl} and cell suspensions \cite{Reasor:2012ey} remain qualitatively consistent across a wide range of viscosity ratios.} The discoid radius $a$ for human $\mathrm{RBC}$ is about $4 \mu \mathrm{m}$. The $\mathrm{RBC}$ membrane in-plane shear elasticity modulus $G \sim 2.5-6 \mu \mathrm{N} / \mathrm{m}$. The deformability of a capsule in the pressure-driven flow is measured by the dimensionless capillary number $C a=\eta \dot{\gamma}_w a / G$. $C a$ is set to be 0.1 for normal RBCs, which corresponds to $\dot{\gamma}_w \sim 100 \mathrm{~s}^{-1}$. {Cell membranes are modeled as isotropic, hyperelastic surfaces characterized by an interfacial shear modulus $G$, incorporating properties of shear elasticity, area dilatation, volume conservation, and bending resistance. The total strain energy of the RBC membrane $S$ is expressed as:
$$
E=\frac{K_B}{2} \int_S\left(2 \kappa_H+c_0\right)^2 d S+\overline{K_B} \int_S \kappa_G d S+\int_S W d S
$$
where $K_B$ and $\overline{K_B}$ are the bending moduli, and $W$ represents the shear strain energy density. Here, $\kappa_H$ and $\kappa_G$ denote the mean and Gaussian curvatures of the surface, respectively, while $c_0 = -2H_0$ represents the spontaneous curvature, with $H_0$ being the mean curvature of the spontaneous shape. The first two terms correspond to the Canham-Helfrich bending energy \cite{canham1970minimum, helfrich1973elastic}, which accounts for the bending resistance of the membrane. The third term represents the shear strain energy stored in the RBC membrane, computed using the Skalak model \cite{skalak1973strain}. More details of the membrane mechanics model, along with validation against experimental observations, are provided in \cite{Sinha:2015wt, Cheng.2023.10.1126/sciadv.adj6423}.}

{In this study, a normal RBC is modeled as a flexible capsule, with a spontaneous biconcave discoidal shape to present shear elasticity and an oblate spheroidal shape for bending elasticity \cite{Sinha:2015wt, EVANS:1972uf}, with a characteristic radius of $a = 4 \, \mu \mathrm{m}$. In SCD, abnormal sickle hemoglobin polymerizes within RBCs upon deoxygenation, forming elongated fibers that disrupt the cellular architecture \cite{bertles1968irreversibly}. This pathological alteration increases membrane stiffness and leads to cellular dehydration, which in turn reduces cell volume. As a result, sickle RBCs are considerably less deformable than normal cells, with some subpopulations irreversibly adopting a sickle-like shape. In this study, we model a sickle cell as a stiff capsule with a curved prolate spheroidal rest shape. The volume of this curved prolate capsule is approximately $20\%$ of that of the normal RBC model.} Since aberrant RBCs in blood disorders are generally much smaller and stiffer than normal RBCs, the interfacial shear modulus $G$ of the aberrant RBCs is assumed to be five times that of normal RBCs, which leads to $C a$ for aberrant RBCs being at most 0.2 times that of $\mathrm{Ca}$ for normal RBCs.  All results are for simulations that have been run to a statistically stationary state.

Fluid-structure interaction is addressed using the immersed boundary method (IBM). Specifically, our model incorporates two types of immersed boundaries: deformable moving cellular membranes and rigid nonmoving vascular walls. The capsule membrane is discretized into $N_{\Delta}$ piecewise flat triangular elements; $N_{\Delta p}=1280$ for normal RBCs, while $N_{\Delta t}=682$ for sickle RBC. Differing values of $N_{\Delta}$ are selected to ensure comparable sizes of triangular elements on both capsules. We employ the ``continuous forcing" IBM and ``direct forcing" IBM methods for the RBC membranes and vessel wall, respectively. The numerical methodology adheres to the approach delineated in previous works \cite{balogh2017computational,mittal2008versatile}. Further details are found in the supplementary information of \CXP{\cite{Cheng.2023.10.1126/sciadv.adj6423}.}

{In this paper, the focus is on understanding the impact of RBC aggregation on hemodynamics in the microcirculation, rather than investigating the detailed mechanisms of RBC aggregation, thus a simple Morse potential is employed to model the interaction forces between RBCs\cite{liu2006rheology}. } In the numerical implementation, intercellular forces are applied between nodes on the triangular mesh representing the RBC membrane surface, so we define below the energy and force per unit area. 
The interaction energy per unit area for nodes separated by a distance $r$, denoted by $\varphi(r)$, is given by the expression: 
\begin{equation}
    \varphi(r)=D_{\mathrm{e}}\left[\mathrm{e}^{2 \beta\left(r_0-r\right)}-2 \mathrm{e}^{\beta\left(r_0-r\right)}\right].
\end{equation}
The corresponding force per unit area, $f(r)$, which describes the interaction forces, is obtained by taking the derivative of the potential: 
\begin{equation}
    f(r)=-\frac{\partial \varphi(r)}{\partial r}=2 D_{\mathrm{e}} \beta\left[\mathrm{e}^{2 \beta\left(r_0-r\right)}-\mathrm{e}^{\beta\left(r_0-r\right)}\right].
\end{equation} 
Here, $r_0$ and $D_{\mathrm{e}}$ represent the zero force distance and surface energy scale per unit area, respectively, while $\beta$ is the inverse interaction length. 
In our simulation, we set  $r_0=0.49 \mu m$ and $\beta=3.84 \mu m^{-1}$\cite{Zhang:2008kr}.  To quantify the strength of aggregation relative to viscous stresses due to flow, we introduce the dimensionless number $K_{\agg}=D_e \beta / \eta \dot{\gamma}_w$. Roughly speaking, at $K_\agg=1$, the forces pulling cells together are balanced by the viscous stress working to separate them. In the context of cell-cell interaction, to improve the computational efficiency, a cut-off length of $4 \mu \mathrm{m}$ is chosen, beyond which the attractive force rapidly diminishes. In the implementation, following finite element discretization of the RBC membrane, a sphere with a diameter equal to the cut-off length is utilized to identify the cell surface within the domain of influence around the node on the cell surface.

In our simulation, we maintain a particle Reynolds number, denoted as $Re_p=\rho \dot{\gamma}_{w} a^{2} / \eta$, at a fixed value of $0.1$. {Although this value of Reynolds number $Re_p$ is high for microcirculation, it represents a compromise between physical accuracy and efficiency. To ensure the robustness of our findings, we performed verifications using $Re_p = 0.05$ and observed no changes in our conclusions.} The fluid is assumed to be incompressible and Newtonian, thereby subjecting the flow to the governing equations of Navier-Stokes and continuity. No-slip boundary conditions are imposed on the tube walls, while periodic boundary conditions are applied in the flow direction. The cell suspension is subjected to unidirectional pressure-driven flow, with the velocity field in the absence of RBCs described by Poiseuille flow within a straight cylindrical tube. In this study, flow is driven by a constant pressure gradient, which is equivalent to fixing the mean wall shear rate at $\dot{\gamma}_w = 2 U_0 / R$, where $U_0$ is the undisturbed centerline velocity. For the curved cylindrical channel, pressure gradients are defined based on an equivalent straight cylindrical channel with the same centerline length and radius. Employing a projection method facilitates the temporal advancement of the velocity field. The cylindrical vessel is positioned within a cuboidal computational domain of $15a \times 10a \times 10a$, utilizing an Eulerian grid with dimensions $150 \times 100 \times 100$. For the curved vessel, a larger cuboidal computational domain of $26a \times 24a \times 10a$ is used, with an Eulerian grid of dimensions $260 \times 240 \times 100$.

\section{Results and discussion}
\subsection{Cylindrical tube}

\begin{figure}[h]
    \centering
    \includegraphics[width = 0.8\linewidth]{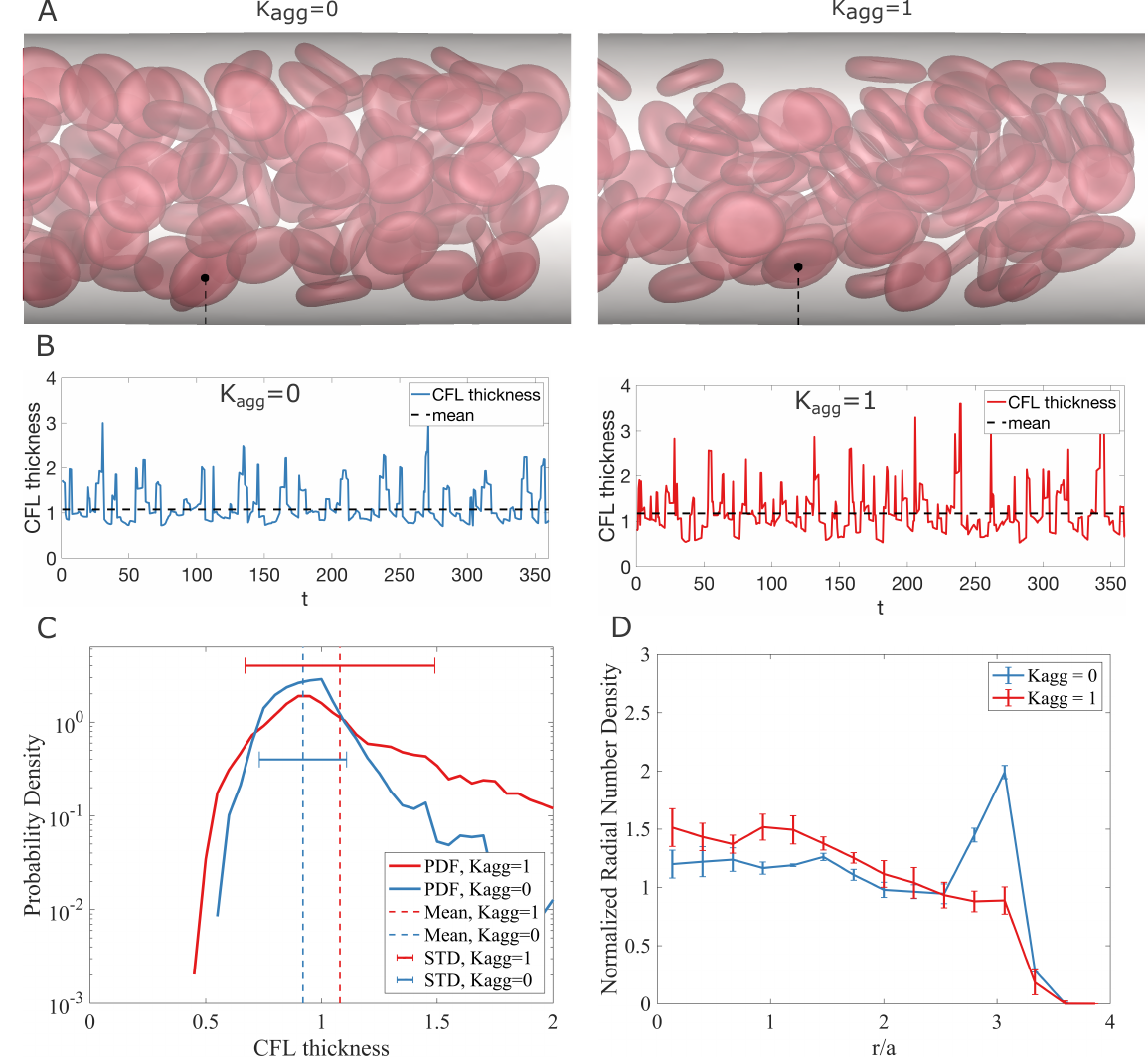}
    \caption{(A) Simulation snapshots of a suspension of RBCs flowing with (left) $K_{agg} = 0$ and (right) $K_{agg} = 1$ through a cylindrical tube with a radius of $16 \mu m$. The direction of flow is from left to right. Hematocrit $\phi=20\%$. Dots and dashed lines represent the position of the cell center-of-mass and the thickness measurements of the nearby cell-free layer, respectively. (B) Temporal evolution of cell-free layer thickness near a fixed point on the cylindrical surface in the cell suspension with (blue) $K_{agg} = 0$ and (red) $K_{agg}=1$. (C) Probability density profile of CFL thickness near the cylindrical surface. (D) Normalized radial number density profile for blood cells in a cylindrical tube. The number density is counted based on the cell center-of-mass position. 
    The profile is normalized such that if the cells are evenly distributed in the tube, then the normalized cell number density is one everywhere.}
    \label{fig:fig1}
\end{figure}

We begin the assessment of aggregation on cell dynamics within a suspension with simulations of RBC suspensions flowing through cylindrical tubes, as presented in Fig.~\ref{fig:fig1}. {Fig.~\ref{fig:fig1}(A) presents simulation snapshots of RBC suspensions with and without aggregation, at a 20\% volume fraction, flowing through a cylindrical channel. In the suspension without RBC aggregation ($K_{\text{agg}} = 0$), RBCs are more evenly distributed within the channel, concentrating primarily in the tube center and forming a cell-free layer along the vessel wall. In contrast, when RBC aggregation occurs ($K_{\text{agg}} = 1$), the cells aggregate into stacks, creating rouleaux-like structures and resulting in large void regions due to the attraction among cells. In the microcirculation, normal RBCs tend to migrate away from the vessel wall, forming a cell-free layer that is governed by the interaction of two cross-stream fluxes: wall-induced migration and hydrodynamic collisions\cite{Kumar:2012ie}. 
The thickness of the cell-free layer varies over time and across different spatial locations near the vessel surface. To quantitatively evaluate the thickness and its variations, we define it as the distance from the center-of-mass position of the cell closest to the wall to the vessel surface, as indicated by the dashed line in Fig.~\ref{fig:fig1} (A). The results from Fig.~\ref{fig:fig1} (A) qualitatively show that in the absence of RBC aggregation, the thickness of the cell-free layer remains relatively uniform. However, with aggregation, the cell-free layer becomes highly heterogeneous, exhibiting substantial fluctuations in its thickness.}

To provide further insight into the CFL, a point is randomly placed on the channel surface, and the thickness of the cell-free layer nearest to this point is measured over time as blood suspension flows. Fig.~\ref{fig:fig1} (B) presents the temporal variation of the cell-free layer thickness near a randomly selected fixed point on the cylindrical vessel surface. It was observed that in the presence of aggregation, the average thickness of the cell-free layer increased, accompanied by a notable amplification in the fluctuation of its thickness. Thus, we observed spatiotemporal non-uniformity in the cell-free layer due to the emergence of cell aggregates, leading to instances where certain cells closely approach the vessel wall. It is important to note that although the CFL thickness is calculated based on the cell center-of-mass position illustrated in Fig.~\ref{fig:fig1}(A), accounting for the cell radius, the actual proximity of the cell to the wall can be significantly close. {In the presence of cell aggregation, the average thickness of the cell-free layer increases, as expected, due to the tendency of cells to attract one another rather than stay close to the vessel wall. However, the variance is also larger, reflecting the formation of cell aggregates.}

{Experiments by Druzak et al. \cite{Druzak.2023.10.1038/s41467-023-37269-3} demonstrated that both the number and size of RBC aggregates increase with rising fibrinogen concentration. In their microfluidic channels, fibrinogen-induced erythrocyte aggregation resulted in glycocalyx reduction and damage to the vascular endothelium.
However, the mechanisms underlying the resulting endothelial damage remain unclear. To investigate this in a computational model, we conducted a statistical analysis of the variations in cell-free layer thickness.
} Fig.~\ref{fig:fig1}(C) displays the probability density curve of the cell-free layer thickness. Notably, the probability density profile for the cell-free layer thickness exhibited a broader distribution in the scenario of cell aggregation. This implies that particularly for very thin cell-free layers, the probability of occurrence is higher when aggregation is present compared to when it is absent, despite the fact that cell aggregation results in an overall increase in the average thickness of the cell-free layer. To describe the statistical distribution of a suspension of cells during flow, the radial distribution of cells within the tube is depicted in Fig.~\ref{fig:fig1}(D). It was observed that in scenarios involving cell aggregation, the peak of cell density adjacent to the cell-free layer decreased, whereas the concentration of cells around the central region of the tube increased. Moreover, in instances of aggregation, there was an augmented deviation in cell distribution, affirming the presence of distribution heterogeneity induced by aggregation.

\begin{figure}[h]
    \centering
    \includegraphics[width = \linewidth]{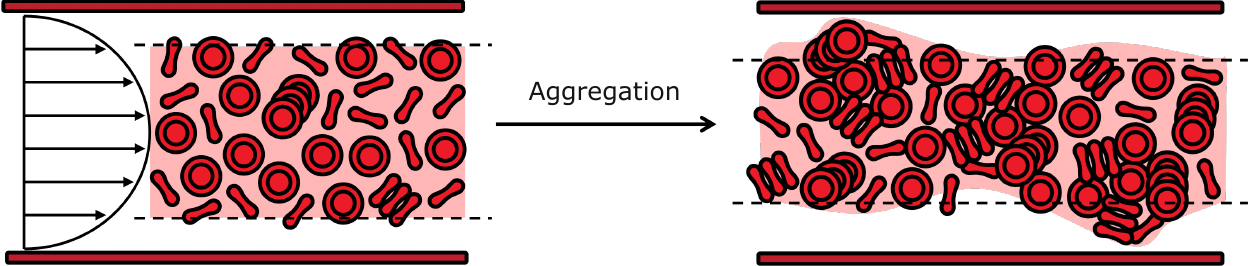}
    \caption{
    Schematic diagram showing the process of RBC aggregation in the blood flow. Nearby red blood cells aggregate into a clump or rouleaux shape as a result of mutual attraction 
    ; As a result, in the presence of cell aggregation, the average thickness of the CFL increased, while CFL fluctuations became more pronounced, and cell aggregates formed very close to the vessel wall.}
    \label{fig:fig2}
\end{figure}

\begin{figure}[h]
    \centering
    \includegraphics[width = \linewidth]{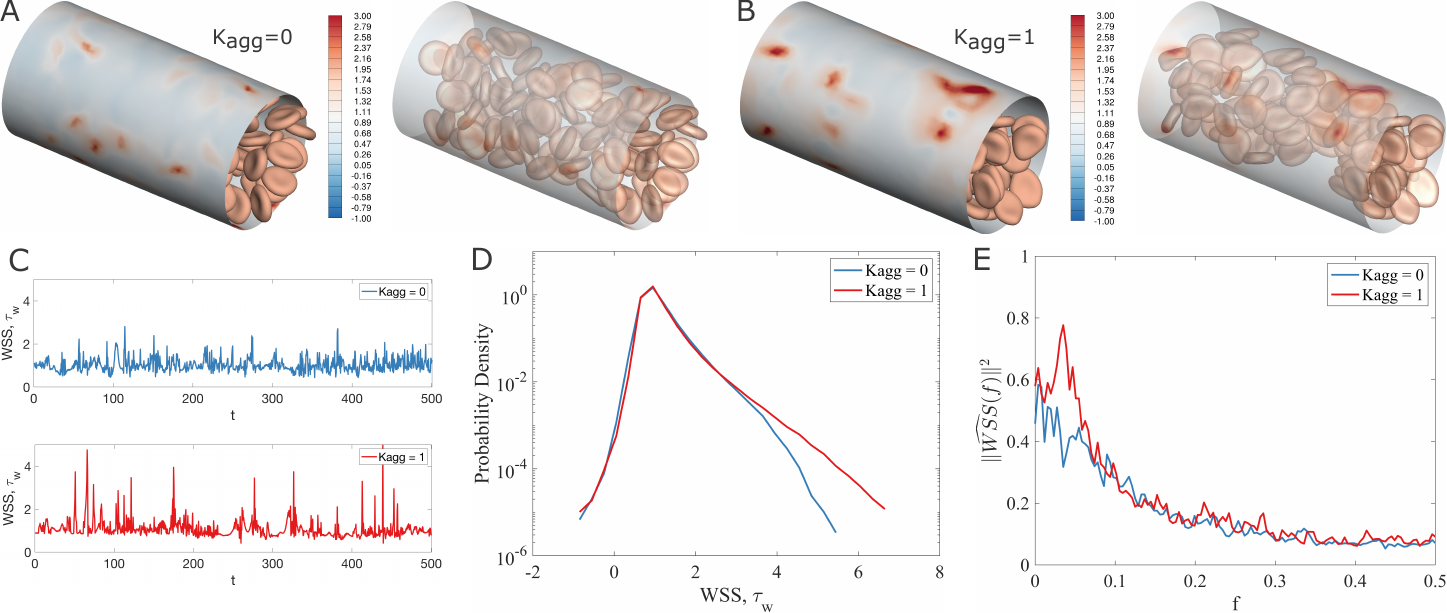}
    \caption{Simulation snapshots of (left) wall shear stress on the cylindrical surface and (right) corresponding transparent views in RBC suspension with (A) $K_{agg} = 0$ and (B) $K_{agg} = 1$. (C) Temporal evolution of wall shear stress at a fixed point on the cylindrical surface in RBC suspension. (D) Probability density profile of wall shear stress on the cylindrical surface. Note that the wall shear stress is dimensionless using $\dot{\gamma}_w \sim 100 \mathrm{~s}^{-1}$. (E) Power spectral density of wall shear stress extracted with the mean, $\tau_w - \bar{\tau}_w$. \CXP{This spectral analysis is performed using a Hamming window.}
    }
    \label{fig:fig3}
\end{figure}

Based on our simulation findings, we propose a hypothesis suggesting that aggregated RBCs form rouleaux or clusters during flow, as shown in Fig.~\ref{fig:fig2}. These clusters approach the blood vessel wall closely, potentially even surpassing the thickness of the cell-free layer in the absence of aggregation. This direct physical contact is posited to induce fluctuations in wall stress, ultimately contributing to the reduction of the glycocalyx and subsequent vascular damage. To study the hydrodynamic impact of cell aggregation on the vascular surface, \MDGrevise{we now characterize the wall shear stress environment experienced without and with aggregation.} Simulation snapshots illustrating wall shear stress are presented in Fig.~\ref{fig:fig3} (A,B). These reveal that local stress fluctuations are more pronounced in RBC suspensions with aggregation than without. Transparent views in Fig.~\ref{fig:fig3} (A) illustrates that RBC aggregates closer to the wall, leading to heightened local stress fluctuations. Conversely, without aggregation, the CFL maintains greater uniformity, thus mitigating stress fluctuations attributed to RBCs being in close proximity to the wall. For better elucidation, Fig.~\ref{fig:fig3} (C) demonstrates the temporal variation of wall stress at a fixed point on the vascular surface. At certain time intervals, the aggregated RBC suspension induces large upward fluctuations in wall stresses due to the proximity of a cell or aggregate. This phenomenon is further illustrated by the probability density profile depicted in Fig.~\ref{fig:fig3} (D), where a substantial disparity in the high-stress tail of the distribution highlights that the likelihood of encountering high wall shear stress in aggregating RBC suspensions is orders of magnitude greater. 

Fig.~\ref{fig:fig3}(E) compares the power spectral density of the wall shear stress (WSS) fluctuations at a point on the wall (averaged over many points) without and with aggregation. {In the present analysis, time is scaled with inverse shear rate so a frequency $f=1$ is a dimensional frequency of $\gammadot$. The results indicate that for $f\gtrsim 0.3$, the spectrum is fairly flat, corresponding to fluctuations due to the overall randomness of the flow in the presence of RBCs and their interactions. }

{Some scaling arguments lead to further insight into the behavior at lower frequencies. We can estimate the streamwise velocity $v$ of cells interacting with the wall to scale as $v\sim \gammadot a$ -- this is the fluid velocity of one particle radius $a$ from the wall. The streamwise flux $j_w$ of cells near the wall satisfies $j_w\sim nv$, where $n$ is the cell number density. This obeys $n\sim\phi a^{-3}$, where $\phi$ is volume fraction (hematocrit). With this result, the streamwise flux $j_w\sim\phi a^2\gammadot$. With these results,  the frequency $f_w$ of cells passing within one radius of a particular point on the wall can be estimated as $f_w\sim j_w a^2$, leading to the final estimate $f_w\sim\phi\gammadot$. We interpret the increase in the PSD for $f\lesssim 0.3$ as the effect of individual cells moving by the sampling point. Some fraction of these will pass close enough to lead to a spike in WSS. For a smooth CFL, ($K_\agg=0$) there is no further structure. In contrast, for $K_\agg=1$, we see a fairly distinct peak centered near $f=0.04$; this probably reflects the average wavelength of the fluctuations in the CFL corresponding to aggregated regions of RBCs passing very close to the wall. This point is corroborated by examination of the $K_\agg=1$ plot on Fig.~\ref{fig:fig3}(C); here there are roughly 18 events with $WSS>2$ over $500$ time units, corresponding to a frequency of $18/500=0.036$, very close to the peak in the PSD.}

% \clearpage
\subsection{Combination of Aggregation and Marginating Sickle Cells}

Recent literature \cite{Cheng.2023.10.1126/sciadv.adj6423} reports that abnormal cells in blood disorders frequently marginate near the vessel wall, generating high wall stresses and potentially causing vessel damage through direct physical contact. However, the influence of cellular aggregation on the segregation process remains insufficiently explored and merits further investigation. To address this, we developed a simplified computational model to study blood flow in the context of SCD, building on our previous studies \cite{Cheng.2023.10.1126/sciadv.adj6423}. {RBC suspensions are modeled as binary mixtures of normal RBCs and aberrant RBCs consisting of 90\% normal RBCs and 10\% irreversibly sickled cells. The introduction of irreversibly sickled cells adds two additional types of cell-cell aggregation interactions: sickle cell-to-sickle cell and sickle cell-to-normal cell. It is known that sickle cells become sticky and can block the vessels. Nevertheless, for simplicity and to isolate the interplay of margination and aggregation, in this study we apply the same cell-cell aggregation model between all cells.} 

\begin{figure}[h]
    \centering
    \includegraphics[width=\linewidth]{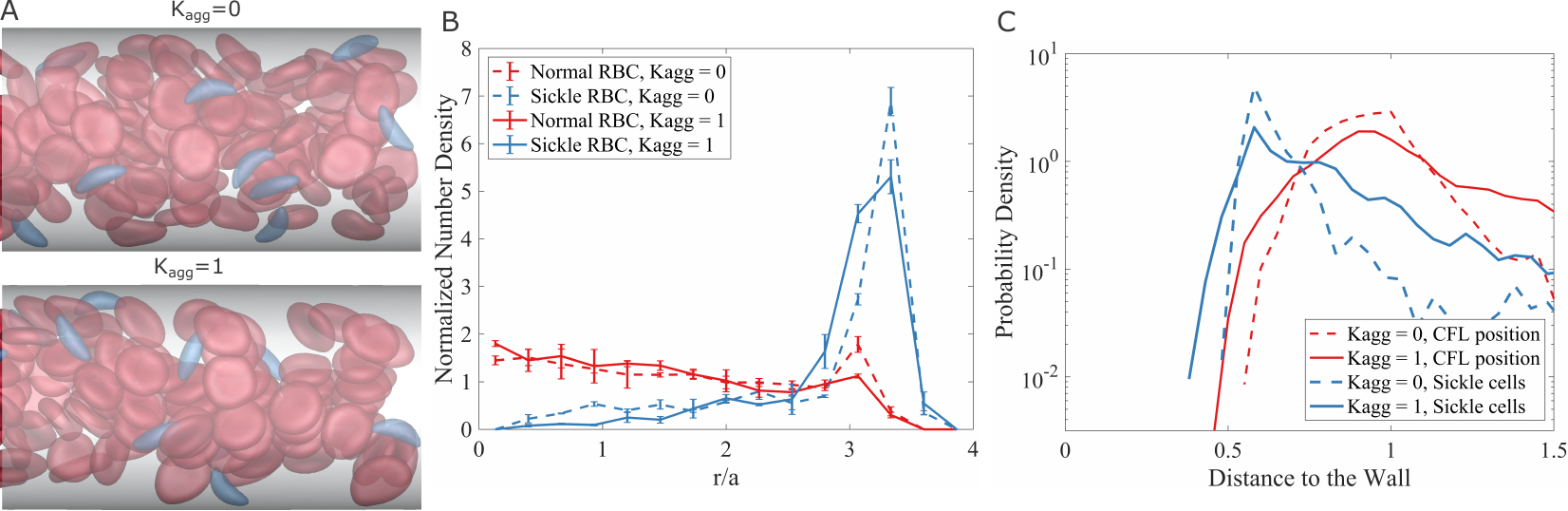}
    \caption{(A) Simulation snapshots of a binary suspension of normal RBCs with sickle cells flowing with (top) $K_\agg=0$ and (bottom) $K_\agg=1$ through a cylindrical tube. Hematocrit is set to $20\%$. (B) The radial cell number density distribution of (red) normal RBCs and (blue) sickle cells within the suspension with (solid line) aggregation and without (dash line) aggregation. (C) Probability density profiling for the distance of (blue) sickle cells and (red) CFL to the walls in the suspension (solid line) with aggregation and (dash line) without aggregation.}
    \label{fig:seg1}
\end{figure}

\begin{figure}[h]
    \centering
    \includegraphics[width=\linewidth]{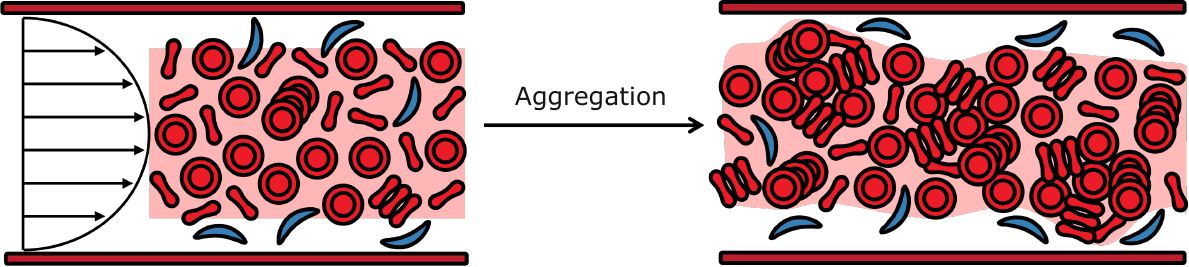}
    \caption{Schematic showing the cell segregation in the aggregation. The perturbation of cell-free layer thickness in the case of RBC aggregation can increase the probability of sickle cell close to the walls.}
    \label{fig:seg2}
\end{figure}

A comparative analysis was conducted for cases with SCD RBCs as illustrated in Fig.~\ref{fig:seg1}. The simulation snapshots in Fig.~\ref{fig:seg1}(A) illustrate the flow of the SCD RBC suspension through a cylindrical channel, comparing scenarios with and without cell aggregation. In the absence of cell aggregation, a uniform cell-free layer is formed by the normal RBC migration, while the stiff sickle cells tend to marginate closer to the vessel wall. With aggregation, however, the mutual attraction between cells results in a more compact distribution of RBCs, while the phenomenon of cell segregation remains remarkable. For quantitative analysis, Fig.~\ref{fig:seg1} (B) presents the radial distributions of normal and sickle cells. In the absence of aggregation, normal cells exhibit a relatively higher concentration peak near the vessel wall. In contrast, with aggregation, normal cells are more concentrated within the center of the channel. When aggregation is absent, sickle cells are primarily positioned near the vessel wall. {However, in the presence of aggregation, the mutual attraction among normal RBCs leads to clusters occupying the channel center, thereby boosting the margination and resulting in a higher proportion of sickle cells within the cell-free layer. In addition, a statistical analysis of cell distribution near the vessel wall is presented in Fig.~\ref{fig:seg1}(C), showing the probability distributions for the locations of the cell-free layer and sickle cells. The majority of sickle cells are positioned outside the cell-free layer. Although cell aggregation induces fluctuations within the cell-free layer, resulting in a broader probability distribution, it also increases the likelihood of sickle cells approaching very close to the vessel wall. This is due to aggregation-induced fluctuations in the CFL thickness, which increase the perturbation in the spatial distribution of sickle cells within CFL, thereby elevating the likelihood of their proximity to the vessel wall, as illustrated in the schematic in Fig.~\ref{fig:seg2}.}

\begin{figure}[h]
    \centering
    \includegraphics[width=\linewidth]{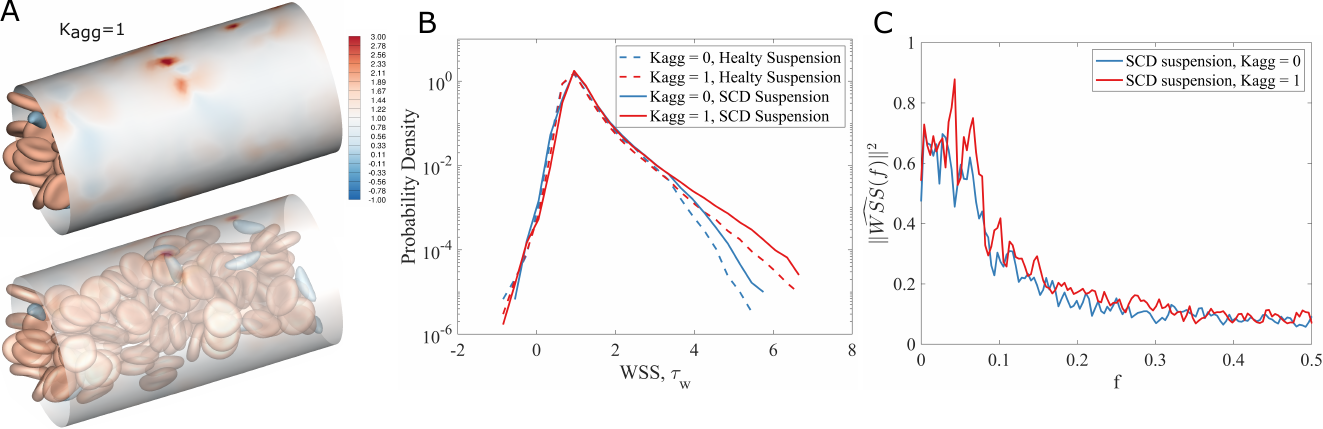}
    \caption{(A) Simulation snapshots of (top) wall shear stress on the cylindrical surface and (bottom) corresponding transparent views in SCD RBC suspension with $K_{agg} = 1$. (B) Probability density profile of wall shear stress $\tau_w$ on the vascular surface. (C) Power spectral density of wall shear stress extracted with the mean, $\tau_w - \bar{\tau}_w$, in SCD suspension. }
    \label{fig:seg3}
\end{figure}

To characterize the hydrodynamic effects of SCD RBC aggregation on the vascular surface, Fig.~\ref{fig:seg3}(A) presents the spatial distribution of wall shear stress. The transparent view on the right reveals that cell aggregation causes the cell-free layer to become highly inhomogeneous. As cells aggregate into clusters, nearby stiff marginated sickle cells approach closely the vessel wall, resulting in large local fluctuations in wall shear stress. Fig.~\ref{fig:seg3}(B) displays the probability distribution of wall shear stress for four different suspensions -- SCD and healthy RBCs, both with and without aggregation. These results indicate that the combination of cell aggregation and the presence of sickle cells leads to enhanced fluctuations in wall shear stress and an increased probability of high wall stress events. {Fig.~\ref{fig:seg3}(C) compares the PSD of wall shear stress fluctuations for SCD RBC suspensions with and without aggregation. Unlike in pure normal RBC suspensions, the case with $K_{\text{agg}} = 1$ exhibits two distinct peaks, centered at around $f = 0.04$ and $f = 0.07$. The peak at $f = 0.04$ observed in the absence of sickle cells persists, with about $10\%$ larger amplitude, while \MDGrevise{in the presence of sickle cells, a higher-frequency peak, at $f = 0.07$, is observed as well. } This high-frequency peak \MDGrevise{may arise from}  the synergistic effect of cell aggregation generating perturbations within the CFL, implying that cell aggregation increases the interaction of sickled cells with the vessel wall.}

\subsection{Effect of Membrane Deformability and Hematocrit in Aggregation}

\begin{figure}[h]
    \centering
    \includegraphics[width = \linewidth]{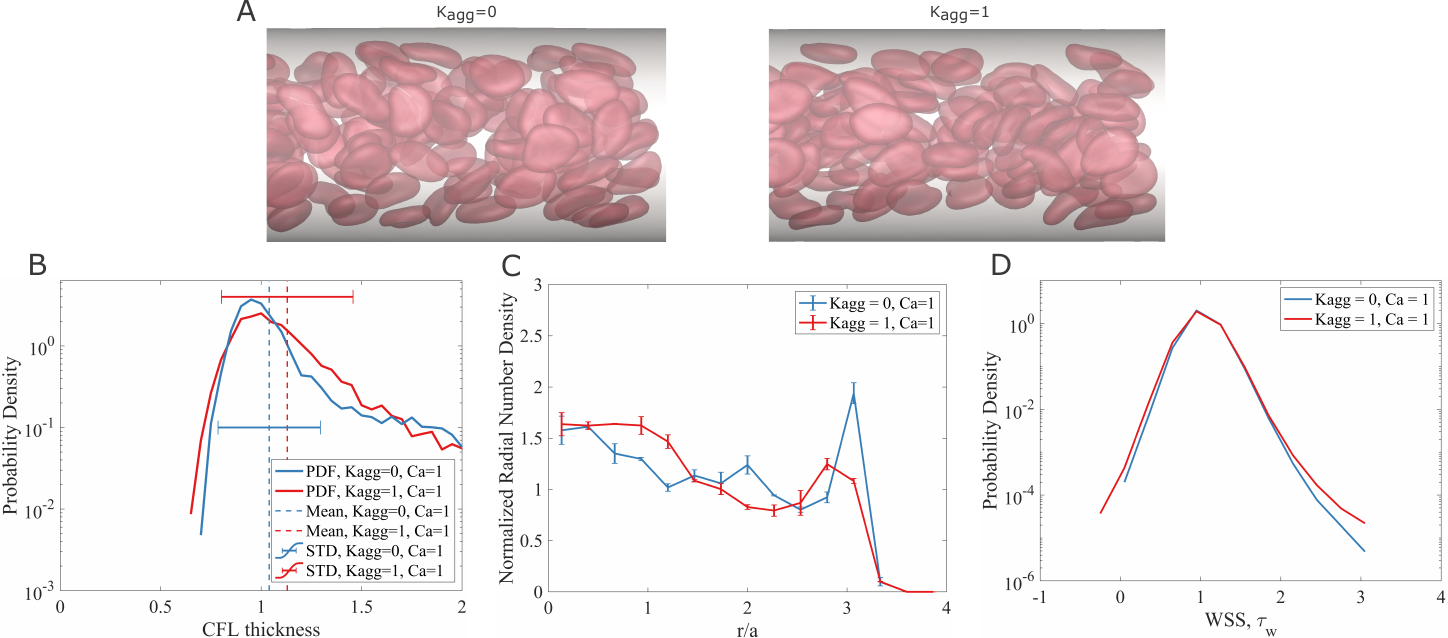}
    \caption{(A) Simulation snapshots of a suspension of RBCs with $\Ca = 1.0$ flowing at (left) $K_{agg} = 0$ and (right) $K_{agg} = 1$ through a cylindrical tube with a radius of $16 \mu m$. Hematocrit is set to $20\%$. (B) Probability density profile of CFL thickness on the cylindrical surface. (C) Normalized radial number density profile for RBCs in a cylindrical tube. (D) Probability density profile for the wall shear stress $\tau_w$ on the vascular surface.}
    \label{fig:soft1}
\end{figure}

{
In the prior sections, the capillary number for the normal RBCs was set to $0.1$, corresponding to a wall shear rate of approximately $100 \, \text{s}^{-1}$. Here, the effect of RBC aggregation at an increased capillary number of $1$ is analyzed, as shown in Fig.~\ref{fig:soft1}.}\MDGrevise{We continue to consider the cases $K_\agg=0$ and $K_\agg=1$. Recalling the definitions $\Ca=\eta\dot{\gamma}a/G$ and $K_\agg=D_e\beta/\eta\dot{\gamma}_w$, increase $\Ca$ by increasing shear rate, to keep $K_\agg$ constant we need to also increase $D_e$. I.e.~we are increasing shear rate, while keeping the balance between aggregation and viscous stress constant. }

{Fig.~\ref{fig:soft1}(A) shows that cells at a higher capillary number ($Ca = 1$) exhibit greater deformations than at $\Ca=0.1$ (as expected), and an apparently more homogeneous cell-free layer compared to those in previous simulations. The latter observation is further supported in Fig.~\ref{fig:soft1}(B), the probability distribution of cell-free layer thickness, demonstrating that at $Ca = 1$, the cell-free layer is generally thicker, and the difference in statistical distribution between the cell-free layer with and without RBC aggregation is relatively small. We attributed this small difference to the stronger wall-induced migration at the higher $\Ca$. Nevertheless, an increased probability of a thinner cell-free layer persists when $K_{\text{agg}} = 1$. Fig.~\ref{fig:soft1}(C) compares the radial distribution of cell number density for cases with and without cell aggregation. In the absence of aggregation, a distinct peak appears near the wall, while in the presence of aggregation, this peak decreases as cells tend to concentrate toward the channel center. Fig.~\ref{fig:soft1}(D) displays the probability distribution of wall shear stress. Although aggregation slightly increases the probability of higher wall shear stress, the effect remains limited; the cell-free layer is much thicker at $\Ca = 1$ than at $\Ca = 0.1$, and aggregation does not substantially influence the wall shear stress distribution. In summary, at high $\Ca$, cell distribution, and wall shear stress statistics are less sensitive to the presence or absence of aggregation, compared with at low $\Ca$.}

\begin{figure}[h]
    \centering
    \includegraphics[width = \linewidth]{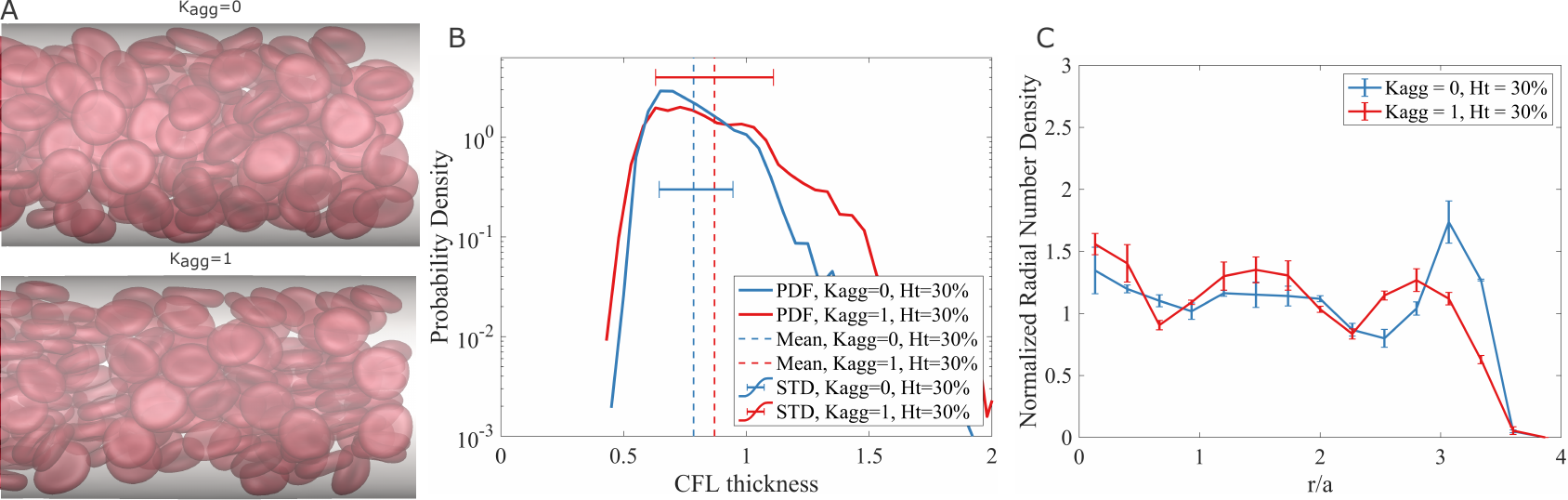}
    \caption{(A) Simulation snapshots of a suspension of RBCs with $Ca = 0.1$ and hematocrit $Ht = 30\%$ flowing at (top) $K_{agg} = 0$ and (bottom) $K_{agg} = 1$ through a cylindrical tube. (B) Probability density profile of CFL thickness on the cylindrical surface. (C) Normalized radial number density profile for blood cells in a cylindrical tube.}
    \label{fig:ht1}
\end{figure}

{We now consider the effect of volume fraction. Hematocrit is not a simple constant in the microcirculation, making it essential to understand the influence of RBC volume fraction on cell aggregation. To address this issue, we now present the distributions and dynamics of RBCs under aggregation conditions, as well as the associated wall stresses, at volume fraction  $\phi = 30\%$. Simulation snapshots in Fig.~\ref{fig:ht1}(A) show and a reduced cell-free layer thickness, as expected at higher $\phi$ \cite{HenriquezRivera:2016wb}. In the presence of cell aggregation, we still observe the formation of rouleaux as well as spatial fluctuations in the cell-free layer; however, unlike at lower hematocrit levels, very large voids of cells do not appear. To further quantify these observations, the probability density distribution of cell-free layer thickness at high hematocrit is presented in Fig.~\ref{fig:ht1}(B). Results indicate that at high hematocrit, the cell-free layer is, on average, thinner than at low hematocrits, as expected due to the increased volume occupied by RBCs. Analysis of the cell-free layer dynamics reveals that, even in the presence of aggregation, instances of a very thin cell-free layer still occur. However, the standard deviation of the cell-free layer thickness is reduced at high hematocrits in both the aggregation and non-aggregation cases.}

\begin{figure}[h]
    \centering
    \includegraphics[width=\linewidth]{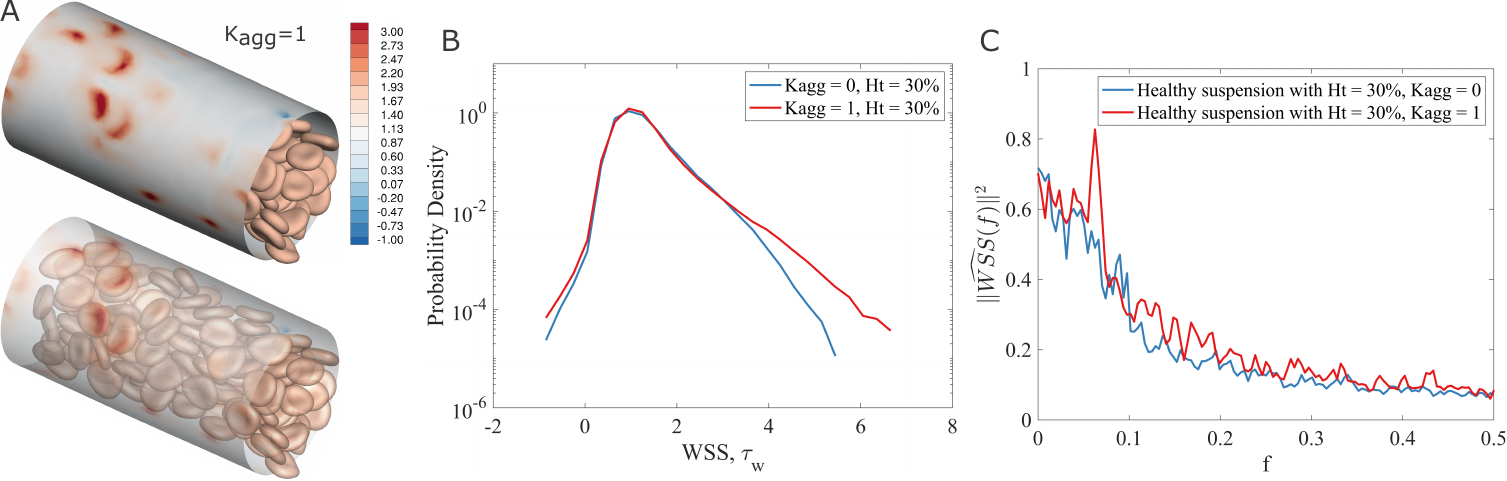}
    \caption{(A) Simulation snapshots of (top) wall
shear stress on the cylindrical surface and (bottom) corresponding transparent views in RBC suspension with intercellular aggregation $K_{agg} = 1$, $Ca = 0.1$, and high hematocrit $Ht = 30\%$. (B) Probability density profile for wall shear stress on the vascular surface in the RBC suspensions with and without aggregation. (C) Power spectrum density of wall shear stress extracted with the mean $\tau_w - \bar{\tau}_w$ in RBC suspension at 30\% hematocrit.}
    \label{fig:ht2}
\end{figure}

{The effect of high hematocrit on vascular wall stress is also examined, as shown in Fig.~\ref{fig:ht2}. In Fig.~\ref{fig:ht2} (A), it is observed that at high hematocrit, the cell-free layer becomes thinner, allowing cells to come into closer proximity with the wall, leading to localized stress fluctuations. Fig.~\ref{fig:ht2} (B) compares cases with and without cell aggregation at high hematocrit, revealing that higher cell concentrations promote aggregation, thereby increasing the likelihood of large stress fluctuations. In Fig.~\ref{fig:ht2} (C), the PSD of wall stress is compared. As at $\phi=0.2$, under conditions of aggregation, the PSD displays a pronounced peak, now at around $f = 0.06$ rather than $f\sim 0.04$, again suggesting that clusters of RBCs contribute to stress fluctuations along the vascular wall. Notably, the $50\%$ increase in the peak frequenct with a $50\%$ increase in $\phi$ is consistent with the physical scaling argument given above, which indicated that $f\sim \phi^1$. 
}

\subsection{Effect of Vascular Geometry: Geometric complexity facilitates RBC clustering}

\begin{figure}[h]
    \centering
    \includegraphics[width = \linewidth]{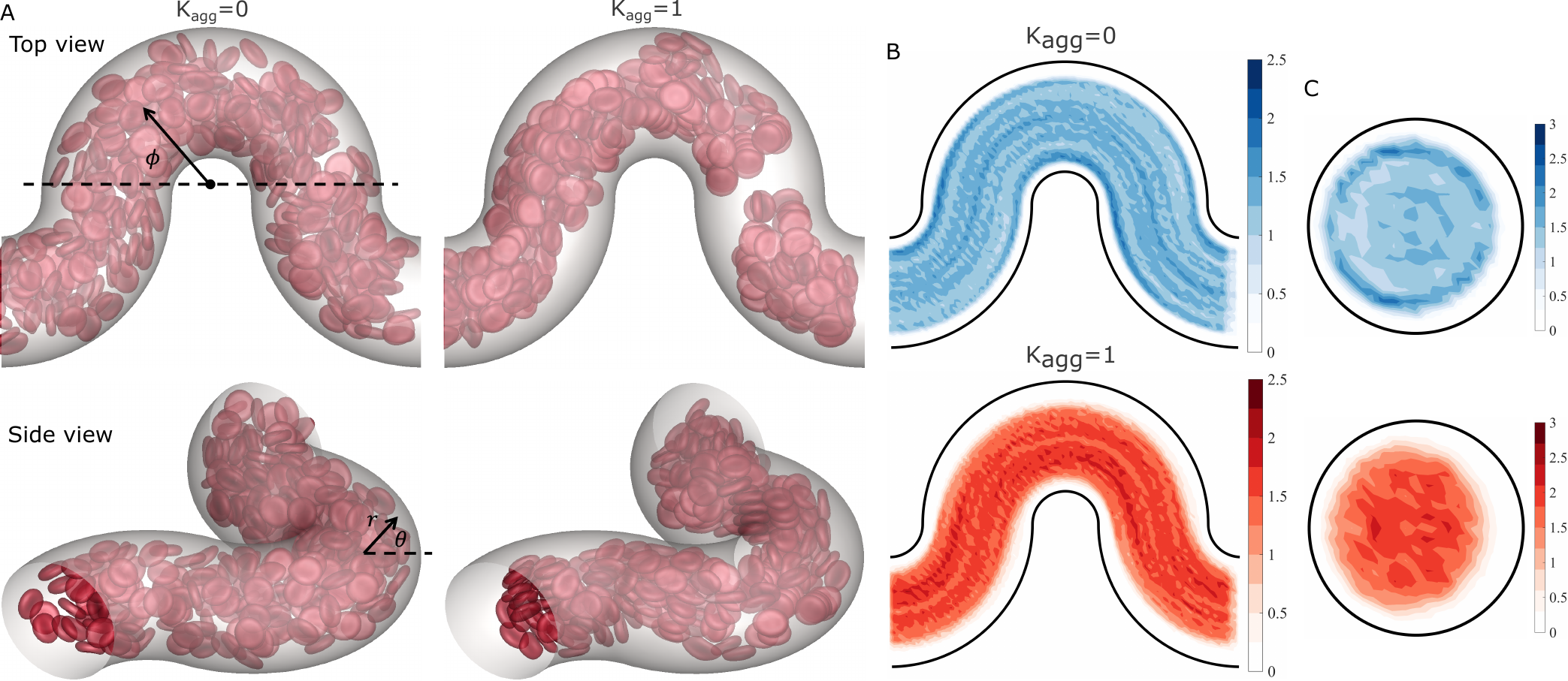}
    \caption{(A) Simulation snapshots of a suspension of RBCs flowing with (left) $K_{agg} = 0$ and (right) $K_{agg} = 1$ through a curved cylindrical tube. Hematocrit, the volume fraction of blood cells, is set to $20\%$. Capillary number $Ca$ is set to $0.1$. The flow direction is from left to right. (B) Center-plane cell number density distribution in a curved channel. (C) Cross-sectional $(0 < \phi < \pi)$ cell number density distribution in a curved channel.}
    \label{fig:curv1}
\end{figure}

In the microcirculation, blood vessels are not simple cylindrical tubes; rather, they exhibit curvature and branching, forming complex vascular networks. Thus, the interaction between vascular geometry and RBC aggregation is studied in this work, as depicted in Fig.~\ref{fig:curv1}. Both aggregation and non-aggregation conditions are considered with Capillary number $\Ca = 0.1$ and hematocrit $20\%$. The simulation snapshots in Fig.~\ref{fig:curv1}(A) reveal that, in the absence of aggregation, RBCs within the curved channel are more uniformly distributed. In contrast, when aggregation occurs, RBCs form rouleaux, aggregates, and even clusters, leading to a markedly heterogeneous spatial distribution with substantial void regions throughout the curved channel. Interestingly, in the case of $K_{\text{agg}} = 1$, the disruption of the "ribbon" structure formed by aggregated RBCs is observed, highlighting the influence of the curved geometry on RBC aggregation and promoting further cell clustering.

\begin{figure}[h]
    \centering
    \includegraphics[width = \linewidth]{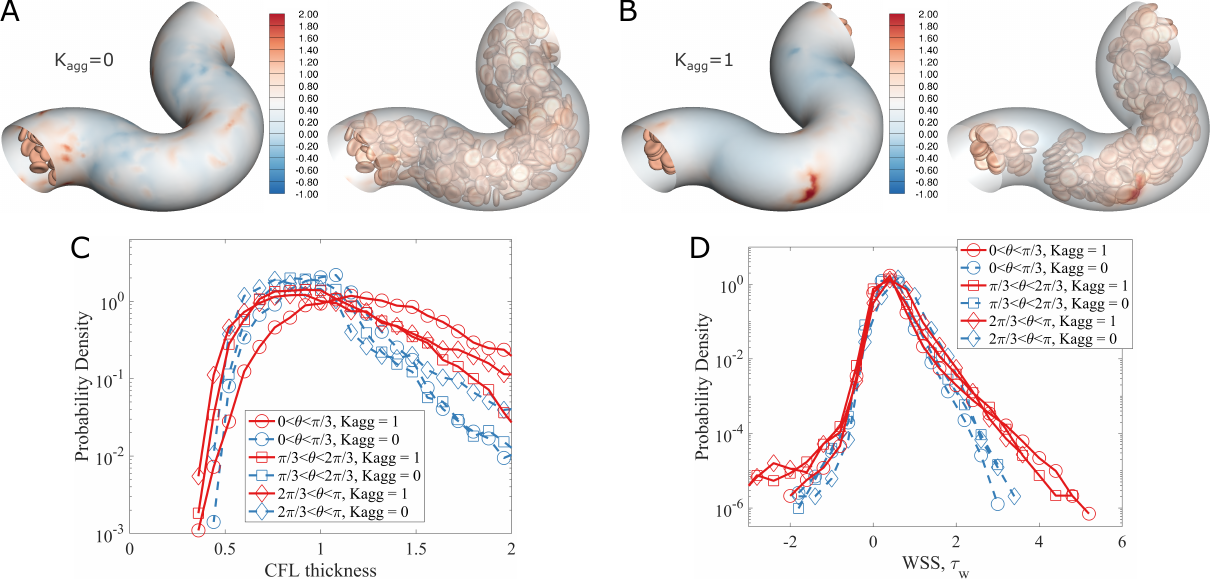}
    \caption{Simulation snapshots of (left) wall shear stress on the curved surface and (right) corresponding transparent views in RBC suspension with (A) $K_{agg} = 0$ and (B) $K_{agg} = 1$. (C) Probability density profile of CFL thickness on the curved surface at different $\theta$. (D) Probability density profile of wall shear stress on the curved surface within different regions, $0 < \theta < \pi/3$ on the outside, $\pi/3 < \theta < 2\pi/3$ in the middle, and  $2\pi/3 < \theta < \pi$ on the inside.}
    \label{fig:curv2}
\end{figure}

\CXP{To further characterize the spatial distribution of RBCs within the curved channel, Figures \ref{fig:curv1}(B) and \ref{fig:curv1}(C) present the cell number density distribution in the center plane and cross-section of the curved channel, respectively. Unlike the straight cylindrical tube, the cell-free layer thickness varies across regions of the curved channel, with the outer side ($\theta < \pi/2$) of the curve exhibiting a thicker cell-free layer compared to the inner side ($\pi/2 < \theta < \pi$). RBC aggregation increases the time-averaged thickness of the cell-free layer in the curved channel while also introducing greater spatial variability. These findings suggest that the geometric complexity of the vessel promotes both instantaneous and time-averaged heterogeneity in the cell-free layer, resulting in larger void regions devoid of cells and encouraging the formation of RBC clusters and clumping during aggregation.}

\CXP{To investigate the hydrodynamic effects of cells on curved vascular surfaces, shear stress distributions along the vascular wall were analyzed, as shown in Fig.~\ref{fig:curv2}(A, B). Notably, Figure \ref{fig:curv2}(B) reveals that RBC aggregation induces large fluctuations in local wall shear stress, particularly along the outer curvature of the vessel, where RBC aggregates accumulate near the wall. Statistical analyses of the cell-free layer thickness and wall shear stress in the inner, middle, and outer regions of the curved channel, depicted in Fig.~\ref{fig:curv2}(C, D), further elucidate these effects. RBC aggregation broadens the probability distribution of the cell-free layer thickness across all regions in Fig.~\ref{fig:curv2}(C), signifying heightened perturbations and an increased likelihood of very thin cell-free layers. The instantaneous distribution of cell-free layer thickness is highly non-uniform, characterized by regions devoid of cells and areas where aggregates approach the wall. In addition, Fig.~\ref{fig:curv2}(D) demonstrates that RBC aggregation elevates the probability of high wall shear stress, particularly on the outer surface of the curved vessel wall, corroborating the localized fluctuations observed in Fig.~\ref{fig:curv2}(B). These results highlight the critical role of RBC aggregation in shaping the hydrodynamic environment of curved vascular.}

\begin{figure}[h]
    \centering
    \includegraphics[width=\linewidth]{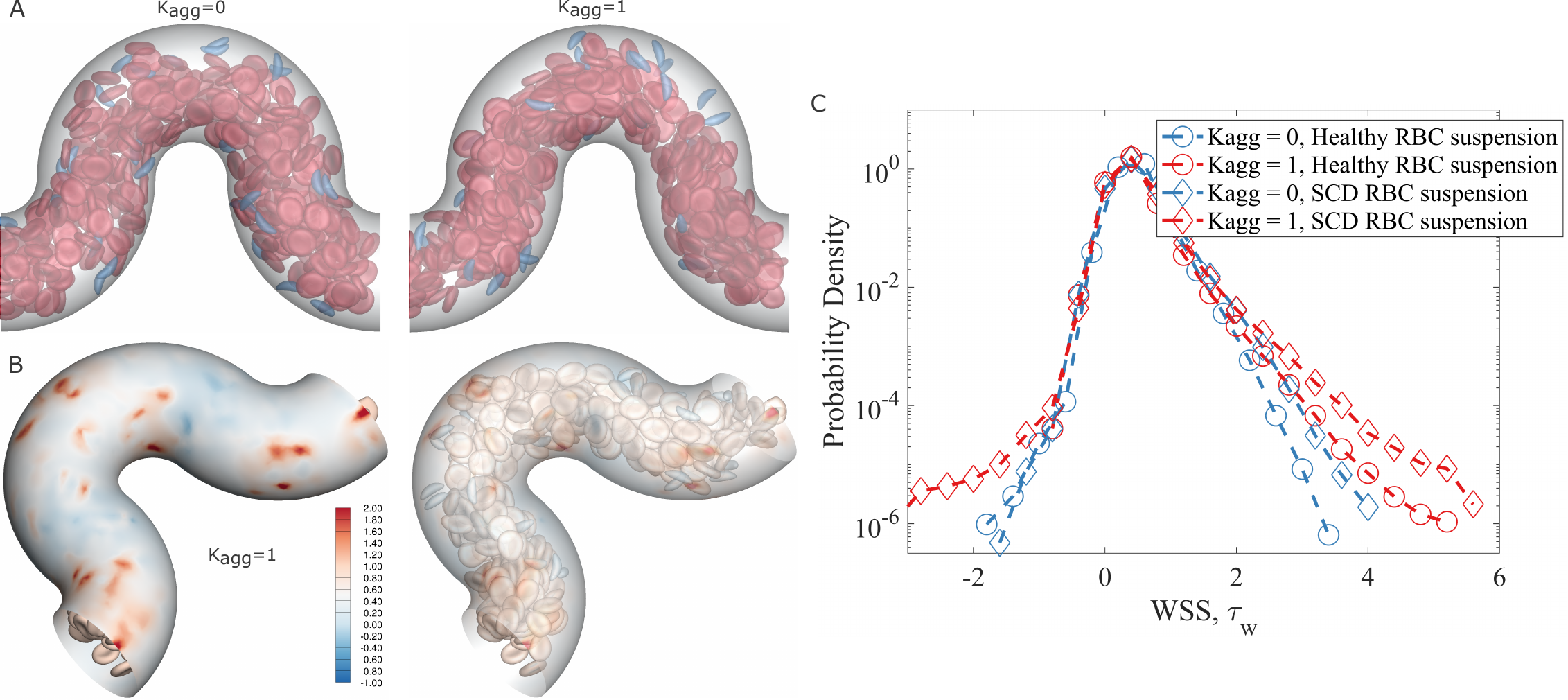}
    \caption{(A) Simulation snapshots of a suspension of SCD RBCs flowing with (left) $K_{agg} = 0$ and (right) $K_{agg} = 1$ through a curved channel. (B) Wall shear stress distribution on the curved vessel surface and its corresponding transparent view in the suspension of SCD RBCs with $K_{agg}=1$. (C) Probability density profile of wall shear stress on the entire curved surface. 
    }
    \label{fig:segcurv1}
\end{figure}

\CXP{The aggregation behavior of our model SCD RBC suspension in a curved channel was also investigated, as shown in Fig.~\ref{fig:segcurv1}(A). Similar phenomena were observed in the curved channel, where RBCs attracted each other, and aggregation led to fluctuations in cell-free layer thickness, boosted the segregation between normal RBCs and sickle cells, and increased the likelihood of these marginating sickle cells being in close proximity to the vessel wall. The distribution of wall shear stress on the curved channel surface in the SCD RBC suspension is shown in Fig.~\ref{fig:segcurv1}(B). Observations from the transparent view revealed that the combination of RBC aggregation and stiff sickle cell margination resulted in large fluctuations in wall shear stress, contributing to vascular damage and inflammation. A statistical analysis of wall shear stress in the curved channel, presented in Fig.~\ref{fig:segcurv1}(C), indicates that the synergistic effects of RBC aggregation and sickle cell margination substantially increase the probability of elevated wall shear stress on the curved vessel surface.}

\section{Conclusion}

This study uses direct simulations to examine the hydrodynamic effects of RBC aggregation on the shear stress environment in small blood vessels.  RBC aggregation leads to a more heterogeneous spatial distribution of RBCs, leading to nonuniformity of the CFL thickness and increasing the probability of large shear stress fluctuations. local wall shear stress fluctuations as cells approach the vessel wall. This effect aligns with experimental findings about increased variation of CFL thickness and endothelial disruption in erythrocyte aggregation, 
\MDGrevise{supporting the hypothesis} that aggregation-induced proximity of RBC clusters to the wall may contribute to glycocalyx reduction and vascular endothelial damage\cite{Druzak.2023.10.1038/s41467-023-37269-3}.

\MDGrevise{With a simple model of sickle cell disease, where there is a small population of smaller and stiffer RBCs, we illustrate the the effects of aggregation on wall shear stress fluctuations are exacerbated by these aberrant cells.}
These cells marginate because of their altered size and stiffness, and in the absence of aggregation, a distinct peak in sickle cell concentration is observed near the wall. Aggregation increases the mean CFL thickness, enhancing segregation and margination, and increases the likelihood of sickle cells approaching the wall due to CFL thickness fluctuations. Accordingly, the presence of both aggregation and aberrant stiff cells leads to a larger probability of large wall shear-stress fluctuations than either aggregation or aberrant cells individually.

Simulations of blood flow through serpentine vessels reveal that vascular geometry significantly influences RBC aggregation and clustering. In the presence of aggregation, RBCs form clusters that lead to an inhomogeneous spatial distribution and increased variability in CFL thickness, as in the straight-tube case. This clustering effect is further accentuated by curvature, which promotes RBC aggregation near the wall, resulting in localized wall shear stress fluctuations. When segmented by region, the inner, middle, and outer areas of the channel exhibit significant perturbations in CFL thickness due to aggregation, increasing the probability of very thin CFL regions and thus elevated wall stresses. In cases involving SCD, aggregation causes sickle cells to marginate closer to the wall, further intensifying wall stress fluctuations and increasing the risk of vascular damage and inflammation.

\section*{Acknowledgments}

Funding: This work was supported by NSF grant CBET-2042221 and ONR grant N00014-18-1-2865 (Vannevar Bush Faculty Fellowship) (M.D.G., X.C. and D.Z.), an American Society of Hematology (ASH) Research Training Award for Fellows (RTAF), NIH National Heart, Lung, and Blood Institute (NHLBI) grant T32HL139443 and Pediatric Loan Repayment Program (LRP) Award L40HL149069, and NIH NHLBI grant R35HL145000 (W.A.L.). The work was performed, in part, at the Georgia Tech Institute for Electronics and Nanotechnology, a member of the National Nanotechnology Coordinated Infrastructure (NCCI), which is supported by the NSF grant ECCS-2025462. This work used the Advanced Cyberinfrastructure Coordination Ecosystem: Services \& Support (ACCESS). In particular, it used the Expanse system at the San Diego Supercomputing Center (SDSC) through allocation MCB190100, and PHY240144. 

Author contributions: M.D.G, W.A.L., and E.I. designed research. X.C. and D.Z. performed research. All authors analyzed data and wrote the paper. 

Declaration of interests: The authors declare that they have no competing interests. 

Data availability: All data needed to evaluate the conclusions in the paper are present in the paper and/or the Supplementary Materials.

\bibliography{bloodMDG, S24proposalrefs, ref, scibib}

\end{document}